\documentclass[english]{scrartcl}
\usepackage[T1]{fontenc}
\usepackage[latin9]{inputenc}
\usepackage{amsmath}
\usepackage{graphicx}
\usepackage[format=plain,font=footnotesize,labelfont=bf]{caption}
\usepackage{color}

\makeatletter

\providecommand{\tabularnewline}{\\}

\usepackage{calc}

\newcommand\mytabspace[1]{%
 \parbox[c][\totalheightof{#1}+2\fboxsep+2\fboxrule][c]{\widthof{#1}}{#1}%
}

\date{}

\makeatother

\usepackage{babel}

\begin{document}

\title{The Role of the Concentration Scale in the Definition of Transfer
Free Energies}

\author{Beate Moeser\textsuperscript{{*}} and Dominik 
Horinek\\
\\
Institute of Physical and Theoretical Chemistry,\\
University of Regensburg,\\
93040 Regensburg, Germany}

\maketitle

\vspace{3cm}

\begin{centering}
Corresponding author:\\

Beate Moeser\\
Institute of Physical and Theoretical Chemistry,\\
University of Regensburg,\\
93040 Regensburg, Germany\\
Phone:+499419434386\\
Beate.Moeser@ur.de\\
\end{centering}

\vspace{3cm}
\textcolor{red}{NOTICE}: this is the author's version of a work that was 
accepted for publication in \textit{Biophysical Chemistry}. Changes 
resulting from the publishing process, such as peer review, editing, 
corrections, structural formatting, and other quality control mechanisms may not 
be reflected in this document. Changes may have been made to this work since it 
was submitted for publication.
A definitive version was subsequently published in \textit{Biophysical 
Chemistry}, Vol.\,196, (2015). DOI 10.1016/j.bpc.2014.09.005

\newpage
\section*{Abstract}
The Gibbs free energy of transferring a solute at infinite dilution
between two solvents quantifies differences in solute-solvent interactions
--- if the transfer takes place at constant molarity of the solute.
Yet, many calculation formulae and measuring instructions that are
commonly used to quantify solute-solvent interactions correspond to
transfer processes in which not the molarity of the solute but its
concentration measured in another concentration scale is constant.
Here, we demonstrate that in this case, not only the change in solute-solvent
interactions is quantified but also the entropic effect of a volume
change during the transfer. Consequently, the ``phenomenon'' which
is known as ``concentration-scale dependence'' of transfer free
energies is simply explained by a volume-entropy effect. Our explanations
are of high importance for the study of cosolvent effects on protein
stability.

\begin{figure*}[b]
\begin{centering}
\includegraphics[width=0.8\textwidth]{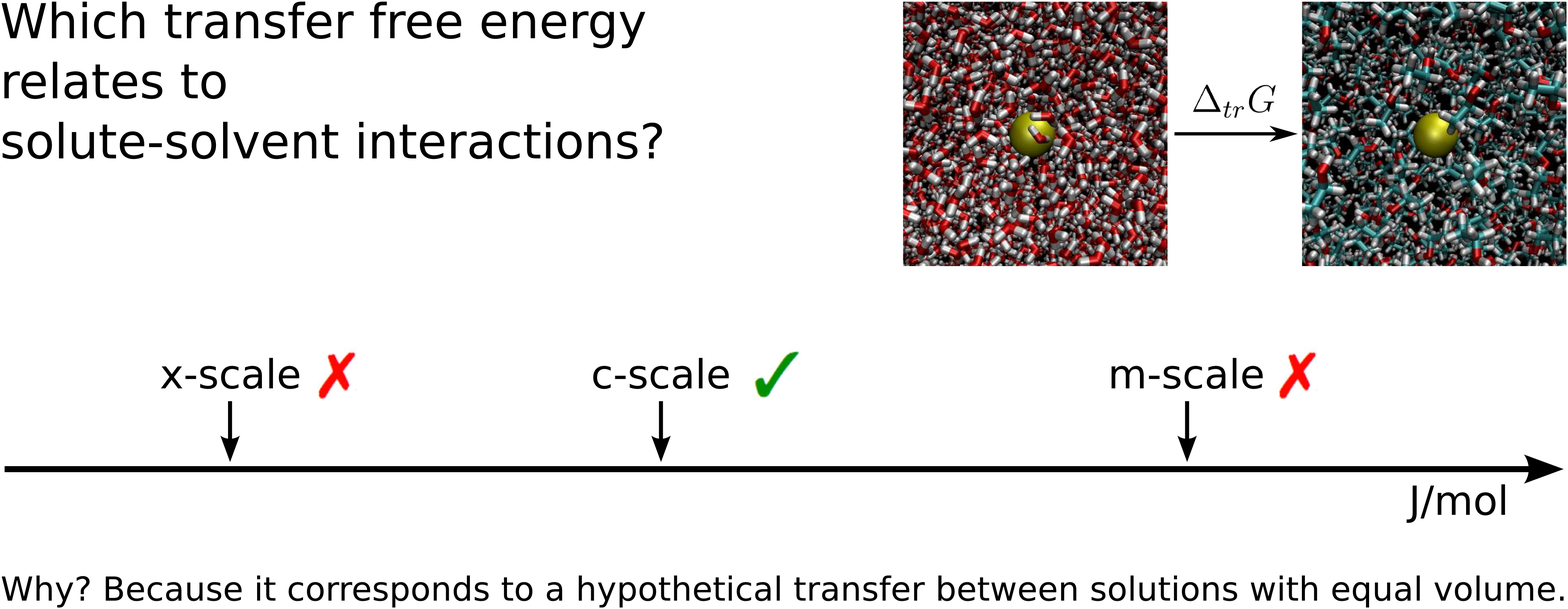}
\par\end{centering}
\end{figure*}

\vspace{3cm}
\noindent Key words: Transfer free energy; Concentration scale; Solute-solvent 
interaction; Cosolvent and osmolyte effects on proteins; Preferential 
interaction; Thermodynamics of solvation.

\newpage

\section{Introduction}

The knowledge of the preference of a solute for one solvent over another
is very important in understanding basic processes in biochemistry,
biology, solution chemistry, and related natural sciences. Moreover,
it is an important prerequisite in the design of products and production
processes in formulation and engineering. Gibbs free energies of transfer
(often abbreviated by ``transfer free energies'' = TFEs) are quantities
that are used to quantify the solvent preferences of solutes. Thus,
many explanatory models and design principles are based on measured
or calculated TFEs. In biochemistry and biology, TFEs are extensively
used in the study of chemical denaturation and renaturation of proteins
or other macromolecules. In the framework of the transfer model, for
example, TFEs can help to unravel which groups of a protein promote
or prevent unfolding in a denaturant or osmolyte \cite{Tanford1964,Moeser2014}.
Moreover, commonly used hydrophobicity scales are based on TFEs \cite{Peters2014}.

The underlying idea in TFE analyses is that a solute favors a solvent
`$b$' over another (`$a$'), if the transfer of the solute at infinite
dilution from solvent `$a$' to solvent `$b$' is favorable. Whether
this is the case can be quantified by the Gibbs free energy of the
transfer (TFE). For a transfer at a constant (infinitely small) concentration
$\xi$ of the solute `$i$', the TFE $\Delta_{tr}G_{i,\xi}^{0}\left(a\rightarrow 
b\right)$
is given by the difference of the solute's standard chemical potentials
in the two solvents
\begin{equation}
\Delta_{tr}G_{i,\xi}^{0}\left(a\rightarrow b\right)=\mu_{i,\xi}^{0}\left(b\right)-\mu_{i,\xi}^{0}\left(a\right),\label{eq:Definition}
\end{equation}
where $\xi$ can be any common concentration scale (e.\,g. molarity,
molality, or mole fraction). For a long time, it was unclear, whether
the change in solute-solvent interactions during a transfer at infinite
dilution is best represented by a transfer at constant molarity or
at constant mole fraction of the solute. Most researchers favored
the mole-fraction scale 
\cite{Nozaki1963,Tanford1973,Tanford1962,Kauzmann1959,Arnett1969,
Cohn1943_Chapter9,Whitney1962} in Eq.\,(\ref{eq:Definition}),
whereas others favored the molarity scale \cite{Robinson1965}. At
the latest when one discovered that the sign of the TFE  can depend
on the used concentration scale \cite{Arnett1969}, it was clear that
the choice of concentration scale is highly relevant.  In 1978, Ben-Naim
was able to resolve the question by means of statistical thermodynamics.
In a very in-depth and insightful article \cite{Ben-Naim1978}, he
showed that only the Gibbs free energy of a transfer at constant molarity
can be interpreted directly in terms of favorable or unfavorable solute-solvent
interaction free energy. While Ben-Naim's paper ``Standard Thermodynamics
of Transfer. Uses and Misuses'' \cite{Ben-Naim1978} by now was cited
more than 300 times, the findings reported therein seem to be rather
unknown nowadays. This is best illustrated by the fact that there
exists a variety of studies, in which TFEs or related quantities are
evaluated at constant mole-fraction (e.\,g. \cite{Gekko1981,Kurhe2011})
or constant (aqua-)molality (e.\,g. \cite{Guinn2011,Diehl2013,Lin1994,Timasheff1993})
and nonetheless are interpreted exclusively in terms of solute-solvent
interactions. The error due to this can be negligible in some cases,
but in others it can be so large that it even affects the classification
of the interactions into the categories ``favorable'' and ``unfavorable''
as we will show later on in Fig.\,\ref{fig:zahlenstrahl}. The fact
that the sign of a TFE can depend on the concentration scale for which
the standard chemical potentials are defined is sometimes called ``concentration-scale
dependence'' of TFEs and is still described as a source of confusion
in the recent literature \cite{Auton2007c,Auton2004}.

In the article at hand, we didactically explain why TFEs calculated
by Eq.\,(\ref{eq:Definition}) only yield the desired information
about solute-solvent interaction free energy if the molarity-scale
standard chemical potentials are used. We start out by recapitulating
that depending on the choice of concentration scale in 
Eq.\,(\ref{eq:Definition}),
the calculated TFE corresponds to a different \emph{hypothetical} transfer
process (insofar as the infinitely small concentration of the transferred
substance is kept constant in a different concentration scale). This
fact is nowadays often not paid attention to. Subsequently, we show
how the TFEs of the different transfer processes can generally be
converted into another and provide a convenient table with explicit
conversion terms. A discussion of the conversion equation reveals
that even in the limit of infinite dilution of
the transferred substance it matters in which concentration scale
the concentration is kept constant.
Considering this, we address the question which of the different transfer
processes at infinite dilution should be used to quantify the solvent
preference of a solute. We show that this is the transfer at constant
molarity and we explain comprehensibly how the TFEs corresponding
to the other processes can be interpreted. Our results also affect
TFE-related quantities as e.\,g. ``chemical potential derivatives''.

\section{Different Transfer Processes at Infinite 
Dilution\label{sec:diffprocesses}}

In the recent literature, the TFE of a solute `$i$' between two solutions
`$a$' and `$b$' is often said to be the difference of the solute's
standard chemical potentials in the two solutions $\mu_{i}^{0}\left(b\right)-\mu_{i}^{0}\left(a\right)$
\cite{Auton2004,Auton2007c}. Even though a standard chemical potential
is only defined in connection with a concentration scale (see section~\ref{sub:ChemPotStandAct}
in the appendix), a concentration scale is often not specified. This
suggests (incorrectly) that the choice of concentration scale for
the standard chemical potential is of no significance. However, here,
we show that depending on the concentration scale of the standard
chemical potentials, a different TFE is obtained that corresponds
to a different transfer process. This was already discussed in the
early days of TFE studies \cite{Nozaki1963,Robinson1965}. 

We start our reasoning by considering a general transfer of a single
solute molecule `$i$' from a solution `$a$' to a solution `$b$'.
The Gibbs free energy associated with the removal or the addition
of a single solute molecule from respectively to a large solution
is by definition the solute's chemical potential $\mu_{i}$ in the
considered solution (respectively the negative thereof in case of
removals). Hence, the Gibbs free energy of the transfer of a single
molecule from a \textit{given} solution `$a$' to a \textit{given}
solution `$b$' is
\begin{equation}
\Delta_{tr}G_{i}\left(a\rightarrow b\right)=\mu_{i}\left(b\right)-\mu_{i}\left(a\right).\label{eq:GeneralTransfer}
\end{equation}
Given that this is a general transfer between two solutions, we realize
that a TFE expressed by the difference of standard chemical potentials
must correspond to a transfer between special solutions --- i.\,e.
solutions for which $\mu_{i}\left(b\right)-\mu_{i}\left(a\right)$
reduces to $\mu_{i}^{0}\left(b\right)-\mu_{i}^{0}\left(a\right)$.
To learn under which conditions this is the case, it is instructive
to express Eq.\,(\ref{eq:GeneralTransfer}) in an arbitrary concentration
scale $\xi$
\footnote{$\xi$ may stand for any of the concentration scales listed in section~\ref{sub:Different-Concentration-Scales}
in the appendix or more generally for any concentration scale that fulfills the 
three
criteria listed in section~\ref{sub:conditions} in the appendix.
}:
\begin{equation}
\Delta_{tr}G_{i}\left(a\rightarrow b\right)=\mu_{i,\xi}^{0}\left(b\right)-\mu_{i,\xi}^{0}\left(a\right)+kT\ln\left(\frac{\gamma_{i,\xi}\left(b\right)\cdot\xi_{i}\left(b\right)}{\gamma_{i,\xi}\left(a\right)\cdot\xi_{i}\left(a\right)}\right).\label{eq:GeneralTransferEvaluatedXi}
\end{equation}
The $\xi_{i}$ describe the concentrations of the solute `$i$' in
the two solutions `$a$' and `$b$' and the $\mu_{i,\xi}^{0}$ and
$\gamma_{i,\xi}$ are the standard chemical potentials and activity
coefficients of the solute in the two solutions in the concentration
scale $\xi$. From Eq.\,(\ref{eq:GeneralTransferEvaluatedXi}), it
is evident that a TFE calculated by the difference of standard chemical
potentials corresponds to a transfer process for which the third term
on the rhs  is zero. This is the case if the solute has the same
infinitely small concentration $\xi_{i}$ in both solutions so that
$\gamma_{i,\xi}\left(b\right)=\gamma_{i,\xi}\left(a\right)=1$ and
$\xi_{i}\left(b\right)/\xi_{i}\left(a\right)=1$. The condition $\xi_{i}\left(b\right)=\xi_{i}\left(a\right)$
is necessary because the third term on the rhs of 
Eq.\,(\ref{eq:GeneralTransferEvaluatedXi})
does not vanish if $\xi_{i}\left(b\right)\approx0$ and $\xi_{i}\left(a\right)\approx0$
but $\xi_{i}\left(b\right)\ne\xi_{i}\left(a\right)$. Hence, depending
on the concentration scale to which the standard chemical potentials
in Eq.\,(\ref{eq:Definition}) belong, a \textit{different TFE} is
calculated that corresponds to a \textit{different transfer process}
because the concentration of the transferred solute is kept constant
in a \textit{different concentration scale}. For the commonly used
concentration scales listed in section~\ref{sub:Different-Concentration-Scales}
in the appendix, this implies concretely: the TFE between a solvent
`$a$' and a solvent `$b$' obtained by Eq.\,(\ref{eq:Definition})
corresponds to the Gibbs free energy of the hypothetical transfer
of the solute
\begin{itemize}
\item from an infinitely large volume of solvent `$a$' to a volume of the
same size of solvent `$b$' if determined by $\mu_{i,c}^{0}\left(b\right)-\mu_{i,c}^{0}\left(a\right)$
(molarity scale),
\item from an infinitely large mass of solvent `$a$' to the same mass of
solvent `$b$' if determined by $\mu_{i,\hat{m}}^{0}\left(b\right)-\mu_{i,\hat{m}}^{0}\left(a\right)$
(molality scale),
\item from an infinitely large number of solvent molecules `$a$' to the
same number of solvent molecules `$b$' if determined by $\mu_{i,x}^{0}\left(b\right)-\mu_{i,x}^{0}\left(a\right)$
(mole-fraction scale),
\item from an infinitely large mass of water to the same mass of water in
a mixed solvent if determined by $\mu_{i,m}^{0}\left(b\right)-\mu_{i,m}^{0}\left(a\right)$
(aquamolality scale).
\end{itemize}
In the measurement of a TFE, the Gibbs free energy of transfer is
not determined by the actual realization of one of the above hypothetical
transfer processes. Instead, the difference of the standard chemical
potentials is determined from experiments at finite concentrations
(as e.\,g. solubility measurements \cite{Auton2007c}). Hence, when
we discuss the above processes in the following, the discussion is
\textit{not about how to transfer} in an experiment, but rather about
\textit{which difference of standard chemical potentials to determine}
(in any kind of suitable experiment).

Unfortunately, neither consistent nor precise terms are in common
use for the description and distinction of different TFEs. In the
following, we try to be precise in the choice of words to avoid misunderstandings.
We will use the word ``TFE'' generally for Gibbs free energies of
any transfer processes of a solute between different solvents. TFEs
that are determined by the difference of ``standard'' chemical potentials
are sometimes called ``standard'' Gibbs free energy of transfer.
Here, we adopt this term and abbreviate them by ``STFE''. Thus,
the different STFEs have in common that they correspond to a transfer
process at constant solute concentration in the limit of infinite
dilution, but they differ in the concentration scale in which the
solute concentration is kept constant. To indicate that the solute
concentration is kept constant in a given concentration scale $\xi$,
we use the term ``$\xi$-scale'' TFE. It is important to note that
in this context, the specification of a concentration scale only defines
the underlying transfer process. The concentration units used in experiments
are unaffected by this and a $\xi$-scale TFE can in principle also
be determined by using a different concentration scale (plus conversion
factors). In symbolic notations in equations, we mark STFEs by the
superscript $0$ (to indicate that we take the difference of two infinite-dilution
standard state chemical potentials) and $\xi$-scale TFEs by the subscript
$\xi$.

\section{Conversion between Standard TFEs}

One might think that during a transfer at infinite dilution, only
changes in solute-solvent interactions can contribute to the TFE and
that at the most the size of this contribution differs between the
different infinite-dilution processes. In this case, the sign of all
STFEs could be used as an indicator for the solvent preference of
the solute. However, this notion is not correct as we clearly show
in the following by a discussion of the conversion terms between the
different STFEs that all correspond to different infinite-dilution
transfer processes. The conversion terms can be derived from the definition
of the standard chemical potential for different concentration scales.
This is done in section~\ref{sec:derivConvmu0} in the appendix and
here we focus on a discussion of the result:

Two STFEs, $\Delta_{tr}G_{i,\xi}^{0}\left(a\rightarrow b\right)$
and $\Delta_{tr}G_{i,\theta}^{0}\left(a\rightarrow b\right)$, that
correspond to a transfer of a solute `$i$' at constant concentration
$\xi$ respectively $\theta$ in the limit of infinite dilution
from a solvent `$a$' to a solvent `$b$' are converted by
\begin{equation}
\Delta_{tr}G_{i,\xi}^{0}\left(a\rightarrow b\right)=\Delta_{tr}G_{i,\theta}^{0}\left(a\rightarrow b\right)-kT\ln\left(\frac{\underset{\theta_{i}\left(b\right)\rightarrow0}{\lim}\left(\frac{\xi_{i}\left(b\right)}{\theta_{i}\left(b\right)}\right)}{\underset{\theta_{i}\left(a\right)\rightarrow0}{\lim}\left(\frac{\xi_{i}\left(a\right)}{\theta_{i}\left(a\right)}\right)}\right).\label{eq:Conversion}
\end{equation}
$\xi_{i}\left(s\right)$ and $\theta_{i}\left(s\right)$ express the
concentration of the solute `$i$' in a solution with solvent `$s$'
in the two different concentration scales $\xi$ and $\theta$.

\begin{table}[h]
\caption{Conversion between different STFEs.\label{tab:ready-to-use} The 
argument
of the logarithm in Eq.\,(\ref{eq:Conversion}) is given for pairs
of the concentration scales defined in 
section~\ref{sub:Different-Concentration-Scales} in the appendix.
In the second column, it is given for general transfers between two
solvents `$a$' and `$b$'. $d_{a}$ and $d_{b}$ are the mass densities
and $M_{a}$ and $M_{b}$ the molar masses of the solvents. In the
third column, the argument of the logarithm is given for the special
case of transfers between water $w$ and a mixed solvent of water
and cosolvent $w+co$. $d_{w}$ and $d_{w+co}$ are the mass densities
of water and the mixed water-cosolvent solution, $M_{w}$ and $M_{co}$
are the molar masses of water and the cosolvent, and $m_{co}$ is
the aquamolality of the cosolvent in the mixed water-cosolvent solution.}
\noindent \centering{}
\begin{tabular}{|c|c|c|}
\hline 
\mytabspace{$\theta,\xi$} & $a\rightarrow b$ & $w\rightarrow 
w+co$\tabularnewline
\hline 
\hline 
$x,c$ & $\frac{d_{b}\cdot M_{a}}{d_{a}\cdot M_{b}}$ & 
\mytabspace{$\frac{d_{w+co}}{d_{w}}\cdot\frac{1+m_{co}M_{w}}{1+m_{co}M_{co}}$}
\tabularnewline
\hline 
$\hat{m},c$ & $\frac{d_{b}}{d_{a}}$ & 
\mytabspace{$\frac{d_{w+co}}{d_{w}}$}\tabularnewline
\hline 
$m,c$ & -- & 
\mytabspace{$\frac{d_{w+co}}{d_{w}}\cdot\frac{1}{1+m_{co}M_{co}}$}
\tabularnewline
\hline 
$\hat{m},x$ & $\frac{M_{b}}{M_{a}}$ & 
\mytabspace{$\frac{1+m_{co}M_{co}}{1+m_{co}M_{w}}$}\tabularnewline
\hline 
$m,x$ & -- & \mytabspace{$\frac{1}{1+m_{co}M_{w}}$}\tabularnewline
\hline 
$m,\hat{m}$ & -- & \mytabspace{$\frac{1}{1+m_{co}M_{co}}$}\tabularnewline
\hline 
\end{tabular}
\end{table}

\begin{figure}[h!]
\vspace{2.6cm}
\begin{centering}
\includegraphics[width=1\textwidth]{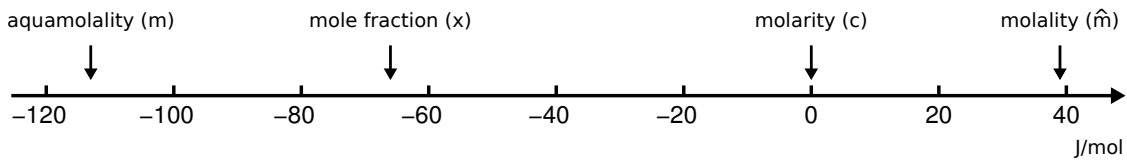}
\par\end{centering}
\caption{Illustration of the difference between different STFEs by means of
the example of transfers between water and a 1\,M urea solution.
The molarity-scale STFE is arbitrarily set to zero. The different
quantities in Tab.\,\ref{tab:ready-to-use} are in the given example:
$d_{w}=0.99707\,\text{kg/L}$, $d_{w+co}=1.01274\,\text{kg/L}$ \cite{Auton2007a},
$m_{co}=1.0497\,\text{mol/kg}$ \cite{Auton2007a}, $M_{co}=60.06\,\text{g/mol}$,
$M_{w}=18.015\,\text{g/mol}$.\protect \\
The molarity-scale STFE of glycine between the two solutions is 17.3\,J/mol
\cite{Moeser2014}, which demonstrates that the illustrated differences
between the different STFEs are not negligible compared to the absolute
values.  \label{fig:zahlenstrahl}\vspace{0.5cm}}
\end{figure}

In Tab.\,\ref{tab:ready-to-use}, we list explicit expressions for
the conversion term evaluated for the complete set of commonly used
concentration scales. It is important to note that the conversion
term is an additive term and not a factor. This implies that if the
STFE is zero for one of the transfer processes it differs from zero
for the others. This is illustrated in Fig.\,\ref{fig:zahlenstrahl}
by means of the example of transfers between water and a 1\,M urea
solution. Hence, we see: It is not possible that the sign of all STFEs
(given by Eq.\,(\ref{eq:Definition})) provides information about
the solvent preference of the solute. Thus, the question arises: Does
any one of the STFEs at all quantify the solute-solvent preference?
If yes, which one? And how are then the other STFEs interpreted?

To our knowledge, Ben-Naim \cite{Ben-Naim1978} was the first to identify
the molarity-scale STFE as the TFE that indeed provides the desired
information about the solvent preferences of solutes, which can also be 
quantified in the framework of the \textit{solvation thermodynamics} introduced 
by him \cite{Ben-Naim1987,Ben-Naim2006}. In the following,
we explain why the molarity-scale STFE has this outstanding interpretation
and explain the physical meaning of the conversion terms.

\section{Interpretation of Standard TFEs\label{sec:Interpretation}}

\begin{figure}
\begin{centering}
\includegraphics[width=7cm]{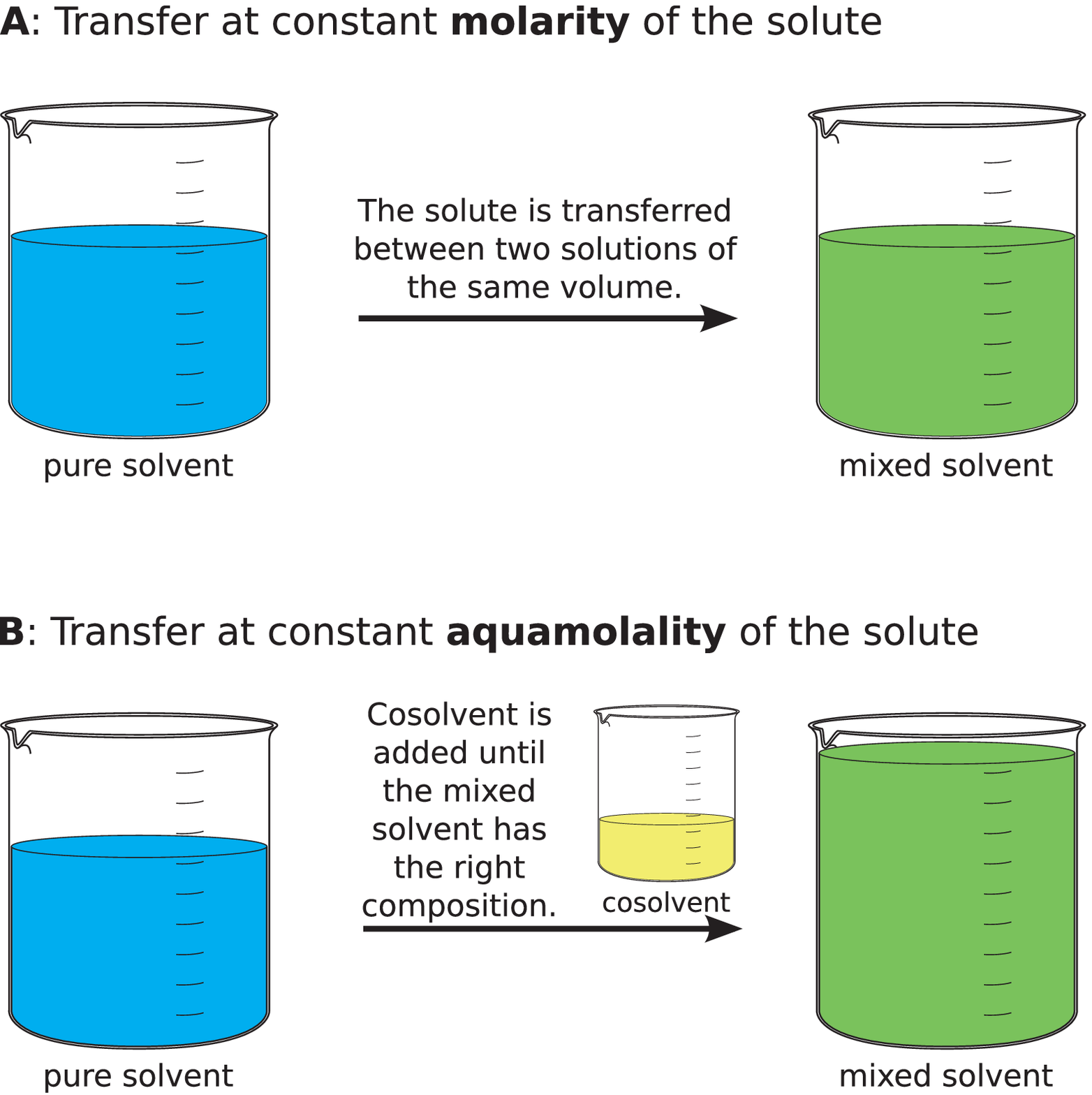}
\par\end{centering}

\caption{\label{fig:processes} Illustration of two different transfer processes
between a pure and a mixed solvent. Panel A shows a transfer whose
change in Gibbs free energy is described by the molarity-scale TFE
(i.\,e. with equal volumes), and panel B shows a transfer whose change
in Gibbs free energy is described by the aquamolality-scale TFE (i.\,e.
with equal masses of the principal solvent water). The schematic drawing
depicts that the two transfer processes are different because the
transfer at constant aquamolality involves a change in accessible
volume. This is entropically favorable so that the aquamolality-scale
TFE is more negative than the molarity-scale TFE for all solutes transferred
between the two depicted solvents.
It is important to note that the figure does not describe how molarity-scale 
and aquamolality-scale TFEs are measured. The figure illustrates the 
essential qualitative difference between the two different hypothetical 
transfer processes whose changes in Gibbs free energy are quantified by 
molarity-scale respectively aquamolality-scale TFEs. See 
section~\ref{sec:diffprocesses} for more explanations.}
\end{figure}

The outstanding interpretation of the molarity-scale STFE can be qualitatively
discussed by means of Fig.\,\ref{fig:processes}. The figure schematically
illustrates the difference of transfer processes between a pure solvent
and a mixed solvent at constant molarity and at constant aquamolality
of the transferred solute:  while the transfer at constant molarity
only involves a change in solvent, the transfer at constant aquamolality
involves in addition to this a change in accessible volume to the
solute. In fact, all possible transfer processes except for those
at constant molarity involve a change in accessible volume. This can
be entropically favorable or unfavorable --- depending on whether
the accessible volume increases or decreases. Yet, TFE studies usually
do not aim at such entropic volume contributions but only at changes
in the solute-solvent interaction free energy  between different
solvents. Hence, we see that the molarity-scale STFE is outstanding
because it corresponds to a hypothetical transfer process that does not 
involve a change in volume and thus exclusively probes changes in interactions 
with the solvents. While a negative molarity-scale STFE implies that the change 
of solvent is favorable, a negative non-molarity-scale STFE implies that the 
change of solvent in combination with the change in volume is favorable. Thus, 
it is fully explainable that the sign of the different STFEs can differ for 
transfers of a given solute between two given solvents.

The validity of this qualitative argument can be proven by means of
statistical thermodynamics and simple mathematics: From the expression
of the standard chemical potential in terms of statistical thermodynamics
as derived by Ben-Naim \cite{Ben-Naim1978} and given in the appendix
(Eqs.\,(\ref{eq:Defmu0}) and (\ref{eq:Defmu0_C})), we get for the
molarity-scale STFE
\footnote{using that the thermal de Broglie wavelength $\Lambda_{i}$ does not
change upon transfer at constant temperature and assuming that the
internal partition function $q_{i}$ does not change either (see 
section~\ref{sub:StatTherm} in the appendix for details).
}
\begin{equation}
\Delta_{tr}G_{i,c}^{0}\left(a\rightarrow b\right)=W\left(i|b^{0}\right)-W\left(i|a^{0}\right).\label{eq:molarityTFE}
\end{equation}
$W\left(i|b^{0}\right)$ and $W\left(i|a^{0}\right)$ are the coupling
works of the solute `$i$' to the pure solvents `$a$' and `$b$'.
Thus, Eq.\,(\ref{eq:molarityTFE}) proves that the STFE of a transfer
at constant molarity directly reflects differences in the solute-solvent
interaction free energy --- expressed here as differences in the coupling
works $W\left(i|b^{0}\right)$ and $W\left(i|a^{0}\right)$. According
to the conversion equation (Eq.\,(\ref{eq:Conversion})), the STFEs
of transfers in which the solute concentration is kept constant in
a scale $\xi$ are given by
\begin{equation}
\Delta_{tr}G_{i,\xi}^{0}\left(a\rightarrow b\right)=W\left(i|b^{0}\right)-W\left(i|a^{0}\right)-kT\ln\left(\frac{\underset{c_{i}\left(b\right)\rightarrow0}{\lim}\left(\frac{\xi_{i}\left(b\right)}{c_{i}\left(b\right)}\right)}{\underset{c_{i}\left(a\right)\rightarrow0}{\lim}\left(\frac{\xi_{i}\left(a\right)}{c_{i}\left(a\right)}\right)}\right).\label{eq:otherTFEs}
\end{equation}
The additional term on the rhs does not depend on the type of solute
`$i$' but only on the concentration-scale conversions in the two
solutions `$a$' and `$b$' between which the transfer takes place.
This fact already excludes that the term is related to solute-solvent
interactions (which moreover are already covered by the first two
summands on the rhs). Instead, the term directly corresponds to the
aforementioned contribution due to a change in volume.
In section~\ref{sec:ProofVolumeIncrease} in the appendix, we prove that
\begin{equation}
-kT\ln\left(\frac{\underset{c_{i}\left(b\right)\rightarrow0}{\lim}\left(\frac{\xi_{i}\left(b\right)}{c_{i}\left(b\right)}\right)}{\underset{c_{i}\left(a\right)\rightarrow0}{\lim}\left(\frac{\xi_{i}\left(a\right)}{c_{i}\left(a\right)}\right)}\right)=-kT\ln\left(\left.\frac{V\left(b\right)}{V\left(a\right)}\right|_{\xi_{i}}\right),\label{eq:IncVol}
\end{equation}
where $\left.\frac{V\left(b\right)}{V\left(a\right)}\right|_{\xi_{i}}$
is the relative increase in volume during a transfer at constant $\xi_{i}$.
This shows that differences in the various STFEs can be 
completely traced back to differences in changes in the accessible volume.
It is important to note that the change in volume in Eq.\,(\ref{eq:IncVol}) is 
due to 
different sizes of the two solutions `$a$' and `$b$' and thus is independent of 
whether the transfer is conducted at constant pressure or at constant volume 
(see section~\ref{sec:NVT} 
in the appendix for further explanations).

With the knowledge of Eq.\,(\ref{eq:IncVol}), we can easily identify the 
conversion terms in
Tab.\,\ref{tab:ready-to-use} with differences in relative increases
in volume between different transfer processes. This is best illustrated
by an example: During a transfer at constant molarity the volume does
not change, but during a transfer at constant molality it changes
because the volume of a solution `$b$' is a factor of $d_{a}/d_{b}$
larger than that of a solution `$a$' with the same mass of another
solvent. Hence, we have $\Delta_{tr}G_{\hat{m}}^{0}=\Delta_{tr}G_{c}^{0}-kT\ln\left(d_{a}/d_{b}\right)$
which agrees with Tab.\,\ref{tab:ready-to-use}.\\

The most relevant implication of the above proof is that non-molarity-scale
STFEs can not be interpreted solely in terms of solute-solvent interaction
free energy. This does not only apply to STFEs as discussed here but
also to related quantities. In protein science, TFEs for transfers
between water and mixed water-cosolvent solutions are for example
often defined by the following equation (e.\,g. \cite{Lin1994,Timasheff1993})
\begin{equation}
\Delta\mu_{tr,2}=\underset{0}{\overset{m_{3}}{\int}}\left(\frac{\partial\mu_{2}}{\partial m_{3}}\right)_{T,P,m_{2}}dm_{3},\label{eq:TFElit}
\end{equation}
where the index $2$ stands for the solute and $3$ for the cosolvent.
$m$ is the concentration in the aquamolality scale. Evaluation of
the integral yields
\begin{equation}
\Delta\mu_{tr,2}=\mu_{2}\left(m_{2},m_{3}\right)-\mu_{2}\left(m_{2},0\right)
\end{equation}
Hence, Eq.\,(\ref{eq:TFElit}) corresponds to the Gibbs free energy
of a transfer of a solute molecule from an aqueous solution to a water-cosolvent
solution that both contain the same aquamolality $m_{2}$ of the solute.
In contrast to the cases discussed before, $m_{2}$ does not need
to be infinitely small. As motivated by Fig.\,\ref{fig:processes}
and proven in the appendix (section~\ref{sec:finiteTFEs}), also
TFEs at constant finite solute concentrations have a contribution
due to volume changes if the transfer is not performed at constant
molarity. Thus, the sign of $\Delta\mu_{tr,2}$ as defined in 
Eq.\,(\ref{eq:TFElit})
can not be interpreted solely in terms of interactions. Similar arguments
apply to the ``preferential interaction parameter'' \cite{Timasheff1993}
which is also called ``chemical potential derivative'' \cite{Guinn2011,Diehl2013,Record2013,Guinn2013a}
\begin{equation}
\mu_{23}=\left(\frac{\partial\mu_{2}}{\partial m_{3}}\right)_{m_{2}}.\label{eq:PrefIntParam}
\end{equation}
This is the integrand of the integral in Eq.\,(\ref{eq:TFElit}).
Under the assumption that the integrand is constant in the considered
interval (cosolvent aquamolality between $0$ and $m_{3}$), the ``$\mu_{23}$
value'' is often determined and interpreted instead of the TFE. If
defined at constant aquamolality as in Eq.\,(\ref{eq:PrefIntParam}),
it also contains a volume contribution so that its sign does not
directly provide information about whether or not the interactions
between the solute and the cosolvent are favorable. Consequently,
if an aquamolality-scale ``$\mu_{23}$ value'' of a molecule is
dissected into contributions of its different surface types (as done
in the solute-partitioning model \cite{Guinn2011,Diehl2013,Record2013,Guinn2013a}),
the entropic volume term is distributed among all surface types present
in the molecule proportionally to the respective areas. Thus, it
affects the thereby determined ``interaction potentials'' of the
surface types, which are meant to quantify interactions between the
surface types and the cosolvent.

\section{Differences in TFEs of Different Solutes}

The change in volume during a transfer at infinite dilution but constant
concentration of the solute in a given concentration scale is independent
of the type of solute that is transferred. It depends only on the
two solutions between which the transfer takes place, which is also
reflected by the fact that the conversion terms in Tab.\,\ref{tab:ready-to-use}
are independent of the solute. Thus, differences in STFEs of different
solutes `$i$' and `$j$' between the same solvents `$a$' and `$b$'
are always independent of the concentration scale in which the solute
concentration is kept constant (provided that it is the same concentration
scale for both solutes). Such differences directly correspond to
the differences in solute-solvent interaction free energy between
the two solutes.

Hence, whenever a study \textit{exclusively} aims at a comparison
of solvent preferences of different solutes, also non-molarity-scale
STFEs can be used (as long as the comparison is accomplished in terms
of differences and not factors). This is for example the case in many
studies that deal with cosolvent effects on protein stability: The
effect of a cosolvent on the folding equilibrium of a protein can
be described by the difference of the TFEs of the native and the denatured
protein structure from water to a cosolvent solution \cite{Tanford1964}.
If $TFE_{D}-TFE_{N}<0$, the transfer of the denatured state to the
cosolvent solution is more favorable than the transfer of the native
state and the protein is denatured by the cosolvent. Analogously,
$TFE_{D}-TFE_{N}>0$ implies stabilization of the protein by the cosolvent.
The quantity $TFE_{D}-TFE_{N}$, which for transfers to a 1\,molar
cosolvent solution corresponds to an $m$ value \cite{Auton2005},
is a difference of TFEs and thus independent of the underlying type
of transfer process.

Also the STFEs of amino acid side chains that are used in the transfer
model for the prediction of cosolvent effects on protein stabilities 
\cite{Nozaki1963,Auton2007c}
are independent of the chosen standard chemical potential difference.
This is because they are defined as the difference of the TFE of the
amino acid with the given side chain and the TFE of glycine, which
does not have a side chain.

\section{Advantageous Concentration Scales in Experiments}

The specification of a concentration in the molarity scale depends
on the temperature, the pressure, and the density of the solution,
whereas mole fractions and (aqua-)molalities are independent of these
influences. Therefore, the preparation of solutions in the latter
concentration scales is much easier than in molarity. Hence, it can
be advantageous to use an experimental setup that determines a non-molarity-scale
STFE and to convert the result to the molarity-scale STFE afterwards
by means of the conversion terms in Tab.\,\ref{tab:ready-to-use}.

Either way --- using the molarity scale right from the start or converting
to molarity scale in the end --- there is no way around using densities
which depend on temperature and pressure. However, during a measurement
of a well-defined and meaningful quantity, the temperature and the
pressure must be fixed anyways --- also if the concentrations are
specified in non-molarity units that are independent of temperature
and pressure.

\section{Summary and Discussion}

We summarize our main messages:
\begin{itemize}
\item The TFE that is determined by the difference of standard chemical
potentials of a concentration scale $\xi$ is the Gibbs free energy
associated with the hypothetical transfer of a solute at (i) infinite
dilution and (ii) constant concentration $\xi$ from one solvent to
another.
\item The sign of the molarity-scale STFE directly provides information
about whether or not the transferred solute prefers one solvent over
another. The sign of non-molarity-scale STFEs can not as readily
be interpreted because these TFEs involve a contribution due to
changes in accessible volume, which is not related to the solute-solvent
interaction free energy.
\item Also TFE-related quantities like ``chemical potential derivatives''
involve contributions due to volume changes if the derivative is not
taken at constant molarity.
\item The conversion terms between different STFEs account for differences
in the increase in volume during the different underlying hypothetical
transfer processes.
\item Differences in STFEs of different solutes (transferred between the
same two solvents) are independent of the underlying transfer process
(as long as the same process is chosen for both of them).
\item It does not matter which concentration scale is used during experiments.
However, if the experiment aims at studying solute-solvent interactions,
the measured TFE has to be converted to a molarity-scale TFE for interpretation.
\end{itemize}
It is obvious that the above insights are of high practical relevance
for the design and analysis of TFE studies for all purposes.

The statistical-thermodynamical ansatz used by Ben-Naim and taken
up by us here is a very straightforward and effective way to find
out which of the different STFEs provides the desired information
about solute-solvent interactions. In 2004, a comprehensive study
\cite{Auton2004} based on a large variety of experiments was published
that among other things also aimed at identifying the STFE that reflects
solute-solvent interactions. In contrast to our reasoning, in that
study the molality-scale TFE is presented as the one that most likely
describes solute-solvent interactions and it is termed ``intrinsic''
TFE. Yet, in the paper it is stressed that --- based on the experimental
evidence --- it can not be excluded that the molarity-scale TFE is
the sought one. Therefore, the authors of the study concluded: ``To
rigorously test the question of preference regarding molal- and molar-based
transfer free energies or whether neither is adequate in all cases,
experiments performed in solvents of widely differing densities will
be required.'' \cite{Auton2004} As explained in the paper at hand,
statistical thermodynamics answers this question without the need
for further experiments: the molarity-scale STFE is the TFE that
quantifies the preference of a solute for one solvent over another.

\section{Acknowledgement}

We thank Sven Essert for helpful discussions and Christoph Hölzl and
Emanuel Schneck for comments on the manuscript.

\newpage
\appendix

\part*{Appendix}

\section{Notation\label{sub:Different-Concentration-Scales}}

The definition and notation for the different concentration scales
used in this article are:
\begin{itemize}
\item molarity $c$ ($\text{\ensuremath{\hat{=}}}$ number density $\rho$):
molecules per volume of the solution.
\item mole fraction $x$: molecules per total number of molecules.
\item molality $\hat{m}$: molecules per mass of solvent (being the mixture
of the principal solvent and the cosolvent in a ternary mixtures).
\item (aqua-)molality $m$: molecules per mass of water.
\end{itemize}
When referring to concentration scales in general, we employ the characters
$\xi$ and $\theta$. Whenever a quantity is only defined in the context
of a given concentration scale (e.\,g. a standard chemical potential
or an activity coefficient), it has an index referring to the concentration
scale.

When different solutions are discussed at the same time, we indicate
in brackets to which solution a given concentration refers.

\section{Standard Chemical Potentials for Different Concentration Scales\label{sec:derivConvmu0}}

\subsection{Representation of the Chemical Potential in Terms of Statistical
Thermodynamics\label{sub:StatTherm}}

We consider a solution of $N_{i}$ solute particles and $N_{j}$ solvent
particles. The chemical potential $\mu_{i}$ of the solute in the
solution at constant pressure and temperature is defined by
\begin{equation}
\mu_{i}=\left(\frac{\partial G}{\partial N_{i}}\right)_{p,T,N_{j}}.
\end{equation}
For sufficiently large solutions, it corresponds to the change in
Gibbs free energy upon the addition of a single solute molecule to
the solution. In \cite{Ben-Naim1978}, Ben-Naim derives an expression
for $\mu_{i}$ by means of statistical thermodynamics that reads
\begin{equation}
\mu_{i}=-kT\ln\left(\left\langle \exp\left(-\frac{\Delta U_{i}\left(r_{0}\right)}{kT}\right)\right\rangle _{0}\right)+kT\ln\left(\frac{\rho_{i}\Lambda_{i}^{3}}{q_{i}}\right).\label{eq:muStatTherm}
\end{equation}
$\rho_{i}$ is the number density (i.\,e. molecules per volume) of
the solute `$i$' in the solution, and $\Lambda_{i}$ and $q_{i}$
are the thermal de Broglie wavelength and the internal partition function
of a molecule of type `$i$'. $\Delta U_{i}\left(r_{0}\right)$ is
the change in system energy upon the addition of a solute molecule
`$i$' at any fixed position $r_{0}$ to the solution at some specific
configuration:
\begin{equation}
\Delta U_{i}\left(r_{0}\right)=U\left(N_{i}+1,N_{j}\right)-U\left(N_{i},N_{j}\right).
\end{equation}
Here, $\left(N_{i},N_{j}\right)$ stands for a specific configuration
of the system and the average $\left\langle \right\rangle _{0}$ in
Eq.\,(\ref{eq:muStatTherm}) is over all configurations of the molecules
in the system except the one that was added.

The first term on the rhs of Eq.\,(\ref{eq:muStatTherm}) is the coupling
work of a molecule of type `$i$' to the solution consisting of $N_{i}$
solute and $N_{j}$ solvent molecules. It accounts for the interactions
between the added particle and the rest of the solution. From now
on we will abbreviate it by $W\left(i|s\right)$ \cite{Ben-Naim1978}
denoting that we couple a molecule of type `$i$' to a solution `$s$',
which is a mixture of molecules of the types `$i$' and `$j$' with
a given composition. If the solute `$i$' is infinitely dilute in
the solution, we write $s^{0}$ so that $W\left(i|s^{0}\right)$ is
the coupling work of the solute to the pure solvent `$j$'. The second
term on the rhs of Eq.\,(\ref{eq:muStatTherm}) corresponds to the
chemical potential that the solute `$i$' had if the solution were
an ideal gas. This term accounts for changes in the Gibbs free energy
$G$ due to internal and translational degrees of freedom of the added
particle and entropic contributions from the indistinguishability
of molecules of type `$i$'.

Using the new notation for the coupling work, we can rewrite 
Eq.\,(\ref{eq:muStatTherm})
\begin{equation}
\mu_{i}=W\left(i|s\right)+kT\ln\left(\frac{\Lambda_{i}^{3}}{q_{i}}\right)+kT\ln\left(\rho_{i}\right),\label{eq:muStatTherm_newNot}
\end{equation}
and see that the chemical potential $\mu_{i}$ of the solute in the
solution depends in two ways on its concentration in the number-density
scale: explicitly through the logarithmic term $kT\ln\left(\rho_{i}\right)$
and implicitly through the dependence of the coupling work $W\left(i|s\right)$
and the internal partition function $q_{i}$ on the solution composition.
For simplicity, we assume throughout the paper, that the internal
partition function of a solute is independent of the surrounding solution.
It is possible to account for solvent effects on the internal partition
function \cite{Ben-Naim1987} and thus to present our results in a
more general form. However, the results themselves do not change if
these effects are taken into account.

\subsection{Representation of the Chemical Potential in Terms of a Standard Chemical
Potential and an Activity Coefficient\label{sub:ChemPotStandAct}}

\subsubsection{Number-Density Concentration Scale}

In the following, we will map the expression in 
Eq.\,(\ref{eq:muStatTherm_newNot})
on a representation of the chemical potential in terms of a standard
chemical potential and an activity coefficient. This representation
is usually used in thermodynamics and here it is needed to derive
the conversion terms between different STFEs. Instead of 
Eq.\,(\ref{eq:muStatTherm_newNot}),
we want to use the following functional form to express $\mu_{i}$
\begin{equation}
\mu_{i}=\mu_{i,\rho}^{0}+kT\ln\left(\rho_{i}\right)+kT\ln\left(\gamma_{i,\rho}\right)\text{\quad with\quad}\underset{\rho_{i}\rightarrow0}{\lim}\left(\gamma_{i,\rho}\right)=1.\label{eq:FuncForm}
\end{equation}
Here, $\mu_{i,\rho}^{0}$ is the standard chemical potential defined
by
\begin{equation}
\mu_{i,\rho}^{0}=\underset{\rho_{i}\rightarrow0}{\lim}\left(\mu_{i}-kT\ln\left(\rho_{i}\right)\right).\label{eq:DefSCP}
\end{equation}
The subscript $\rho$ denotes that it is defined in the number-density
scale. Evaluation of Eq.\,(\ref{eq:DefSCP}) with help of 
Eq.\,(\ref{eq:muStatTherm_newNot})
shows that $\mu_{i,\rho}^{0}$ accounts for the coupling work $W\left(i|s^{0}\right)$
at infinite dilution of the solute as well as for its internal partition
function and parts of the translational partition function $q_{trans}=\frac{V}{\Lambda_{i}^{3}}$
\begin{equation}
\mu_{i,\rho}^{0}=W\left(i|s^{0}\right)+kT\ln\left(\frac{\Lambda_{i}^{3}}{q_{i}}\right).\label{eq:Defmu0}
\end{equation}
$\gamma_{i,\rho}$ is the activity coefficient of the number-density
scale. It is a function of the solute concentration $\rho_{i}$. Insertion
of Eq.\,(\ref{eq:Defmu0}) in Eq.\,(\ref{eq:FuncForm}) and comparison
with Eq.\,(\ref{eq:muStatTherm_newNot}) reveals that the term 
$kT\ln\left(\gamma_{i,\rho}\right)$
accounts for the dependence of the coupling work on the solute concentration,
i.\,e.
\begin{equation}
kT\ln\left(\gamma_{i,\rho}\right)=W\left(i|s\right)-W\left(i|s^{0}\right).\label{eq:Defgamma}
\end{equation}
Hence, we have shown that with the relations (\ref{eq:Defmu0}) and
(\ref{eq:Defgamma}), it is possible to express the chemical potential
$\mu_{i}$ in terms of a standard chemical potential and an activity
coefficient.

\subsubsection{Other Concentration Scales\label{sub:conditions}}

The molarity scale (moles per volume) is essentially the same concentration
scale as the number-density scale (molecules per volume). It differs
from the latter only by a concentration-independent factor $N_{A}$,
the Avogadro constant. Therefore, we do not discriminate between the
two in the main text of the article and call both of them molarity.
We get for the molarity scale in analogy to Eq.\,(\ref{eq:FuncForm})
\begin{eqnarray}
\mu_{i,c} & = & \mu_{i,c}^{0}+kT\ln\left(\gamma_{i,c}\cdot c_{i}\right)\text{\quad with\quad}\underset{c_{i}\rightarrow0}{\lim}\left(\gamma_{i,c}\right)=1\label{eq:FuncFormC}\\
\text{with\quad}\mu_{i,c}^{0} & = & \mu_{i,\rho}^{0}+kT\ln\left(N_{A}\right)\label{eq:Defmu0_C}\\
\text{and\quad}\gamma_{i,c} & = & \gamma_{i,\rho}.\label{eq:GammaP=00003DGammaC}
\end{eqnarray}

Most other concentration scales differ from the molarity and the number-density
scale by a factor that depends on the concentration of the solute.
Nevertheless, $\mu_{i}$ can be expressed in the same functional form
as in Eqs.\,(\ref{eq:FuncForm}) and~(\ref{eq:FuncFormC}) for all
concentration scales $\xi$ that fulfill the following three criteria:
\begin{enumerate}
\item The zero point is the same as in the molarity scale: $\xi\left(c=0\right)=0$.
\item $\xi\left(c\right)$ is strictly monotonic.
\item $\xi\left(c\right)$ is continuous.
\end{enumerate}
If these criteria hold, the zeroth-order term of the Taylor expansion
of $\xi\left(c\right)$ in the point $c=0$ is zero and the first-order
term exists:
\begin{equation}
\xi\left(c\right)=\left.\frac{\partial\xi}{\partial c}\right|_{c=0}\cdot c+\mathcal{O}\left(c^{2}\right).\label{eq:Taylor}
\end{equation}
All concentration scales listed in 
section~\ref{sub:Different-Concentration-Scales} of the appendix
fulfill these criteria and thus $\xi$ may represent any of them.
To express $\mu_{i}$ in the concentration scale $\xi$ by
\begin{equation}
\mu_{i}=\mu_{i,\xi}^{0}+kT\ln\left(\gamma_{i,\xi}\cdot\xi_{i}\right)\text{\quad with\quad}\underset{\xi_{i}\rightarrow0}{\lim}\left(\gamma_{i,\xi}\right)=1,\label{eq:FuncForm_Xi}
\end{equation}
we have to find expressions for $\mu_{i,\xi}^{0}$ and $\gamma_{i,\xi}$
that guarantee that the right-hand sides of Eqs.\,(\ref{eq:FuncForm_Xi})
and (\ref{eq:FuncFormC}) are identical for all concentrations. From
this requirement follows
\begin{equation}
\mu_{i,c}^{0}-\mu_{i,\xi}^{0}=kT\ln\left(\frac{\xi_{i}}{c_{i}}\right)+kT\ln\left(\frac{\gamma_{i,\xi}}{\gamma_{i,c}}\right).\label{eq:Detmu0Xi}
\end{equation}
The lhs of Eq.\,(\ref{eq:Detmu0Xi}) is per definition independent
of the solute concentration. Thus, we can determine $\mu_{i,c}^{0}-\mu_{i,\xi}^{0}$
by evaluating the rhs in the limit of infinite dilution
\begin{equation}
\mu_{i,c}^{0}-\mu_{i,\xi}^{0}=kT\ln\left(\underset{c_{i}\rightarrow0}{\lim}\left(\frac{\xi_{i}}{c_{i}}\right)\right)+kT\ln\left(\underset{c_{i}\rightarrow0}{\lim}\left(\frac{\gamma_{i,\xi}}{\gamma_{i,c}}\right)\right).\label{eq:Detmu0Xi_2}
\end{equation}
Due to the first criterion above, $\gamma_{i,\xi}$ and $\gamma_{i,c}$
are both unity in the considered limit so that the second term on
the rhs of Eq.\,(\ref{eq:Detmu0Xi_2}) vanishes. According to 
Eq.\,(\ref{eq:Taylor}),
the limit $\underset{c_{i}\rightarrow0}{\lim}\left(\frac{\xi_{i}}{c_{i}}\right)$
exists and equals $\left.\frac{\partial\xi_{i}}{\partial c_{i}}\right|_{c_{i}=0}$.
Hence, we can identify $\mu_{i,\xi}^{0}$ as
\begin{equation}
\mu_{i,\xi}^{0}=\mu_{i,c}^{0}-kT\ln\left(\underset{c_{i}\rightarrow0}{\lim}\left(\frac{\xi_{i}}{c_{i}}\right)\right)=\mu_{i,c}^{0}-kT\ln\left(\left.\frac{\partial\xi_{i}}{\partial c_{i}}\right|_{c_{i}=0}\right).\label{eq:relmu0}
\end{equation}
Inserting this back into Eq.\,(\ref{eq:Detmu0Xi}), we get an expression
for $\gamma_{i,\xi}$:
\begin{eqnarray}
\gamma_{i,\xi} & = & \underset{c_{i}\rightarrow0}{\lim}\left(\frac{\xi_{i}}{c_{i}}\right)\cdot\frac{c_{i}}{\xi_{i}}\cdot\gamma_{i,c},\label{eq:convgamma_1}\\
 & = & \left.\frac{\partial\xi_{i}}{\partial 
c_{i}}\right|_{c_{i}=0}\cdot\frac{c_{i}}{\xi_{i}}\cdot\gamma_{i,c}.
\label{eq:convgamma}
\end{eqnarray}
Thus, with the relations~(\ref{eq:relmu0})--(\ref{eq:convgamma}),
it is possible to express the chemical potential in any concentration
scale $\xi$ that fulfills the above criteria in terms of a standard
chemical potential and an activity coefficient.

\subsubsection{General Conversions between Standard Chemical Potentials}

The relations~(\ref{eq:relmu0})--(\ref{eq:convgamma}) can easily
be generalized to relations between the chemical potentials and activity
coefficients of any two concentration scales that fulfill the above
listed criteria. This is because if the criteria hold for two scales
$\xi$ and $\theta$, then they also hold between them. Thus, having
shown that $\mu_{i,\xi}^{0}$ and $\gamma_{i,\xi}$ exist for the
concentration scale $\xi$, we can repeat the derivations~(\ref{eq:Detmu0Xi})--(\ref{eq:convgamma})
with the concentration scales $\theta$ and $\xi$ instead of $\xi$
and $c$, and obtain the general conversion equations:

Standard chemical potentials of any two concentration scales $\xi$
and $\theta$ that fulfill the above listed criteria are converted
by
\begin{equation}
\mu_{i,\xi}^{0}=\mu_{i,\theta}^{0}-kT\ln\left(\underset{\theta_{i}\rightarrow0}{\lim}\left(\frac{\xi_{i}}{\theta_{i}}\right)\right)=\mu_{i,\theta}^{0}-kT\ln\left(\left.\frac{\partial\xi_{i}}{\partial\theta_{i}}\right|_{\theta_{i}=0}\right).\label{eq:Convmu0}
\end{equation}
The corresponding activity coefficients are converted by
\begin{equation}
\gamma_{i,\xi}=\underset{\theta_{i}\rightarrow0}{\lim}\left(\frac{\xi_{i}}{\theta_{i}}\right)\cdot\frac{\theta_{i}}{\xi_{i}}\cdot\gamma_{i,\theta}.
\end{equation}

\subsection{Conversion between Standard TFEs}

A STFE $\Delta_{tr}G_{i,\xi}^{0}\left(a\rightarrow b\right)$ corresponding
to a transfer process at constant $\xi$ is given by the difference
of the $\xi$-scale standard chemical potentials of the solute `$i$'
in the two solutions `$a$' and `$b$' between which it is transferred
\begin{equation}
\Delta_{tr}G_{i,\xi}^{0}\left(a\rightarrow b\right)=\mu_{i,\xi}^{0}\left(b\right)-\mu_{i,\xi}^{0}\left(a\right).
\end{equation}
Thus, the difference between a $\xi$-scale STFE $\Delta_{tr}G_{i,\xi}^{0}\left(a\rightarrow b\right)$
and a $\theta$-scale STFE\\ $\Delta_{tr}G_{i,\theta}^{0}\left(a\rightarrow 
b\right)$
follows directly from Eq.\,(\ref{eq:Convmu0}) and can be expressed
by Eq.\,(\ref{eq:Conversion}).

\section{Proof that the Conversion Term Corresponds to the Relative Increase
in Accessible Volume\label{sec:ProofVolumeIncrease}}

Here, we prove the validity of Eq.\,(\ref{eq:IncVol}). We start out
by recasting the argument of the logarithm in the conversion term
on the lhs of Eq.\,(\ref{eq:IncVol}) in a different form. In the
course of this, we employ several times that the three conditions
for $\xi\left(c\right)$ listed in section~\ref{sub:conditions} in the appendix
hold:

\begin{align}
\frac{\underset{c_{i}\left(b\right)\rightarrow0}{\lim}\left(\frac{\xi_{i}
\left(b\right)}{c_{i}\left(b\right)}\right)}{\underset{c_{i}
\left(a\right)\rightarrow0}{\lim}\left(\frac{\xi_{i}\left(a\right)}{c_{i}
\left(a\right)}\right)} & = 
\frac{\underset{c_{i}\left(a\right)\rightarrow0}{\lim}\left(\frac{c_{i}
\left(a\right)}{\xi_{i}\left(a\right)}\right)}{\underset{c_{i}
\left(b\right)\rightarrow0}{\lim}\left(\frac{c_{i}\left(b\right)}{\xi_{i}
\left(b\right)}\right)}\\
 & =
\frac{\left.\frac{\partial\left(\frac{N_{i}\left(a\right)}{V\left(a\right)}
\right)}{\partial\xi_{i}\left(a\right)}\right|_{c_{i}\left(a\right)=\xi_{i}
\left(a\right)=0}}{\left.\frac{\partial\left(\frac{N_{i}\left(b\right)}{
V\left(b\right)}\right)}{\partial\xi_{i}\left(b\right)}\right|_{c_{i}
\left(b\right)=\xi_{i}\left(b\right)=0}}\\
 & = \frac{V\left(b\right)}{V\left(a\right)}\cdot\frac{\left(\frac{\partial 
N_{i}\left(a\right)}{\partial\xi_{i}\left(a\right)}-\frac{N_{i}\left(a\right)}{
V\left(a\right)}\cdot\frac{\partial 
V\left(a\right)}{\partial\xi_{i}\left(a\right)}\right)_{c_{i}\left(a\right)=\xi_
{i}\left(a\right)=0}}{\left(\frac{\partial 
N_{i}\left(b\right)}{\partial\xi_{i}\left(b\right)}-\frac{N_{i}\left(b\right)}{
V\left(b\right)}\cdot\frac{\partial 
V\left(b\right)}{\partial\xi_{i}\left(b\right)}\right)_{c_{i}\left(a\right)=\xi_
{i}\left(a\right)=0}}\displaybreak[2]\\
 & = \frac{V\left(b\right)}{V\left(a\right)}\cdot\frac{\left.\frac{\partial 
N_{i}\left(a\right)}{\partial\xi_{i}\left(a\right)}\right|_{c_{i}
\left(a\right)=\xi_{i}\left(a\right)=0}}{\left.\frac{\partial 
N_{i}\left(b\right)}{\partial\xi_{i}\left(b\right)}\right|_{c_{i}
\left(b\right)=\xi_{i}\left(b\right)=0}}.\label{eq:InsertBack}
\end{align}
In the last conversion we used that $N_{i}/V=c_{i}\cdot N_{A}=0$
in the considered limit. The numerator and the denominator of the second
factor in Eq.\,(\ref{eq:InsertBack}) are equal, which is proven in
the following by taking into account that the particle numbers $N_{i}\left(a\right)$
and $N_{i}\left(b\right)$ are equal because
all solute particles that are in `$a$'
are transferred to 
`$b$'.
At infinite dilution, this is a single molecule.
Moreover, the concentrations
$\xi_{i}\left(a\right)$ and $\xi_{i}\left(b\right)$ are equal because
of the construction of the transfer process, in which $\xi_{i}$ is
the very property kept constant. Therefore, the ratios of $N$ and
$\xi$ are equal as well
\begin{equation}
\frac{N_{i}\left(a\right)}{\xi_{i}\left(a\right)}=\frac{N_{i}\left(b\right)}{
\xi_{i}\left(b\right)}
\end{equation}
Inserting the Taylor expansion of $N\left(\xi\right)$ in the dilute
limit $\xi\rightarrow0$

\begin{equation}
N\left(\xi\right)=\left.\frac{\partial N}{\partial\xi}\right|_{\xi=0}\cdot\xi+\mathcal{O}\left(\xi^{2}\right)
\end{equation}
yields the desired relation
\begin{equation}
\left.\frac{\partial N_{i}\left(b\right)}{\partial\xi_{i}\left(b\right)}\right|_{\xi_{i}\left(b\right)=0}=\left.\frac{\partial N_{i}\left(a\right)}{\partial\xi_{i}\left(a\right)}\right|_{\xi_{i}\left(a\right)=0},
\end{equation}
which was to be proven. Inserting this back into Eq.\,(\ref{eq:InsertBack}),
finally yields
\begin{equation}
\frac{\underset{c_{i}\left(b\right)\rightarrow0}{\lim}\left(\frac{\xi_{i}\left(b\right)}{c_{i}\left(b\right)}\right)}{\underset{c_{i}\left(a\right)\rightarrow0}{\lim}\left(\frac{\xi_{i}\left(a\right)}{c_{i}\left(a\right)}\right)}=\frac{V\left(b\right)}{V\left(a\right)}.
\end{equation}
$V\left(b\right)/V\left(a\right)$ is the relative increase in volume
during the considered transfer process at constant $\xi$.\\

Evaluation of $\partial N_i/\partial \xi_i|_{\xi_i=0}$ in concrete terms 
reveals that it is the quantity that needs to be the same in both solutions 
`$a$' and `$b$' so that the transfer takes place at constant $\xi$: it is the 
volume for $\xi=c$, the solvent mass for $\xi=\hat{m}$, the mass of the 
principal solvent (water) for $\xi=m$, and the number of molecules for $\xi=x$.

\section{TFEs at Constant Finite Concentrations\label{sec:finiteTFEs}}

According to Eq.\,(\ref{eq:GeneralTransferEvaluatedXi}), the Gibbs
free energy $\Delta_{tr}G_{i,\xi}^{f}$ of the transfer of a solute
molecule between two solutions with the same solute concentration
$\xi$ that is not infinitely small but finite is given by

\begin{equation}
\Delta_{tr}G_{i,\xi}^{f}\left(a\rightarrow b\right)=\mu_{i,\xi}^{0}\left(b\right)-\mu_{i,\xi}^{0}\left(a\right)-kT\ln\left(\frac{\gamma_{i,\xi}\left(b\right)}{\gamma_{i,\xi}\left(a\right)}\right).
\end{equation}
With Eqs.\,(\ref{eq:otherTFEs}), (\ref{eq:convgamma_1}), 
(\ref{eq:GammaP=00003DGammaC}),
and (\ref{eq:Defgamma}) this reduces to
\begin{equation}
\Delta_{tr}G_{i,\xi}^{f}\left(a\rightarrow b\right)=W\left(i|b\right)-W\left(i|a\right)-kT\ln\left(\frac{\frac{\xi_{i}\left(b\right)}{c_{i}\left(b\right)}}{\frac{\xi_{i}\left(a\right)}{c_{i}\left(a\right)}}\right).\label{eq:generalizationFinite}
\end{equation}
 Eq.\,(\ref{eq:generalizationFinite}) corresponds to a generalization
of Eq.\,(\ref{eq:otherTFEs}) to transfers at finite concentrations.
Also here, the argument of the logarithm can be identified with the
relative increase in volume during a transfer at constant $\xi_{i}$
and constant
\footnote{Usually, also at finite concentration, transfer processes with a constant
number of solute particles are considered. If however, a transfer
process is considered in which a solute is transferred between two
solutions with $\xi_{i}\left(a\right)=\xi_{i}\left(b\right)$ but
$N_{i}\left(a\right)\neq N_{i}\left(b\right)$, the argument of the
logarithm in Eq.\,(\ref{eq:generalizationFinite}) does not reflect
the relative increase in volume during the considered process. Instead,
it reflects the relative increase in volume during the corresponding
constant-particle-number transfer between the two considered solutions
with composition $\xi_{i}\left(a\right)=\xi_{i}\left(b\right)$.
} particle number $N_{i}$. Hence, we have
\begin{equation}
\Delta_{tr}G_{i,\xi}^{f}\left(a\rightarrow b\right)=W\left(i|b\right)-W\left(i|a\right)-kT\ln\left(\left.\frac{V\left(b\right)}{V\left(a\right)}\right|_{\xi_{i}}\right),\label{eq:ConstFiniteEq}
\end{equation}
and see that the TFE at finite concentration also consists of an interaction
term and a term due to an increase in volume (which is only zero for
transfers at constant molarity). The interaction term describes the
difference in coupling work of a solute molecule `$i$' to the two
given solutions with different solvents and a finite solute concentration
$\xi_{i}$. A range of transfers that all start from the same solution
`$a$' and keep the solute concentration constant (in different concentration
scales) in general results in different solution compositions of the
second solution `$b$'. Thus, the interaction term in 
Eq.\,(\ref{eq:ConstFiniteEq})
also depends on the concentration scale in which the solute concentration
is kept constant. This is different from the case of STFEs 
(Eq.\,(\ref{eq:otherTFEs}))
where the solutions to which we couple are always pure solvents so
that the interaction term is the same for all STFEs (of a solute transferred
between the same two solvents).

\section{The Meaning of Constant Pressure and Constant Volume\label{sec:NVT}}

Finally, we elucidate the meaning of ``constant pressure'' and ``constant
volume'' conditions during a transfer process. The (hypothetical)
process discussed in this paper is the transfer of a single solute
molecule `$i$' from a solution `$a$' to a solution `$b$', which
comprises two steps: the removal of the solute from system `$a$'
and its insertion into system `$b$'. Each of the two steps can be
performed either at constant volumes $V\left(a\right)$ and $V\left(b\right)$,
or at constant pressure, in which case the two volumes are NPT \textit{ensemble
averages}. As nicely discussed by Ben-Naim \cite{Ben-Naim1987}, the chemical 
potential is identical in both ensembles for
macroscopically large systems. The crucial difference is that at constant
pressure the chemical potential relates to a change in Gibbs free
energy, at constant volume it relates to a change in Helmholtz free
energy. Hence, the Helmholtz free energy $\Delta_{tr}F_{i,\xi}^{0}$
of a transfer at constant volume and the Gibbs free energy 
$\Delta_{tr}G_{i,\xi}^{0}$
of a transfer at constant pressure between the same two systems are
identical. Both are given as
$\mu_{i,\xi}^{0}\left(b\right)-\mu_{i,\xi}^{0}\left(a\right)$ (compare 
Eq.\,(\ref{eq:Definition})).

Finally, we stress that the transfer under NVT conditions must not
be confused with a hypothetical transfer process, in which the volumes
$V\left(a\right)$ and $V\left(b\right)$ are equal. A process with
$V\left(a\right)=V\left(b\right)$ can either be performed at constant
pressure or at constant volume. The condition $V\left(a\right)=V\left(b\right)$
only fixes the scale for which this hypothetical process is characteristic
(the molarity scale) and is no contradiction to the fact that Gibbs
free energies are measured at constant pressure and not at constant
volume.


\newpage
\begin{thebibliography}{10}
\bibitem{Tanford1964} Tanford C.  {Isothermal unfolding of globular Proteins in aqueous urea   solutions}.  {J Am Chem Soc} 1964; 86:2050--2059.
\bibitem{Moeser2014} Moeser B, Horinek D.  {Unified description of urea denaturation: backbone and side chains   contribute equally in the transfer model.}  {J Phys Chem B} 2014; 118:107--114.
\bibitem{Peters2014} Peters C, Elofsson A. {Why is the biological hydrophobicity scale more accurate than earlier experimental hydrophobicity scales?} {Proteins} 2014; doi: 10.1002/prot.24582
\bibitem{Nozaki1963} Nozaki Y, Tanford C.  {The solubility of amino acids and related compounds in aqueous urea   solutions}.  {J Biol Chem} 1963; 238:4074--4081.
\bibitem{Tanford1973} Tanford C.  {The hydrophobic effect: formation of micelles and biological   membranes}. New York:  John Wiley \& Sons, 1973. 200 p.
\bibitem{Tanford1962} Tanford C.  {Contribution of hydrophobic interactions to the stability of the   globular conformation of Proteins}.  {J Am Chem Soc} 1962; 84:4240--4247.
\bibitem{Kauzmann1959} Kauzmann W.  {Some factors in the interpretation of protein denaturation}.  In: Anfinsen Jr. CB, Anson ML, Bailey K, Edsall JT, editors. {Adv Protein Chem, Volume XIV}, New York: Academic Press; 1959. p 1--63.
\bibitem{Arnett1969} Arnett EM, McKelvey DR.  {Solvent isotope effect on thermodynamics of nonreacting solutes}.  In: Coetzee JF, Ritchie CD, editors. {Solute-solvent   interactions}. New York: Marcel Dekker; 1969. p 343--398.
\bibitem{Cohn1943_Chapter9} Cohn EJ, Edsall JT.  {Interactions between organic solvents and dipolar ions estimated   from solubility ratios}.  In: Cohn EJ, Edsall JT, editors. {Proteins, amino   acids, and peptides as ions and dipolar ions}. New York: Hafner Publishing Group; 1965. p 196--216.
\bibitem{Whitney1962} Whitney PL, Tanford C.  {Solubility of amino acids in aqueous urea solutions and its   implications for the denaturation of Proteins by urea}.  {J Biol Chem} 1962; 237:PC1735--PC1737.
\bibitem{Robinson1965} Robinson DR, Jencks WP.  {The effect of compounds of the urea-guanidinium class on the   activity coefficient of acetyltetraglycine ethyl ester and related   compounds}.  {J Am Chem Soc} 1965; 1432:2462--2470.
\bibitem{Ben-Naim1978} Ben-Naim A.  {Standard thermodynamics of transfer. Uses and misuses}.  {J Phys Chem} 1978; 82:792--803.
\bibitem{Gekko1981} Gekko K.  {Mechanism of polyol-induced protein stabilization: solubility of   amino acids and diglycine in aqueous polyol solutions.}  {J Biochem} 1981; 90:1633--1641.
\bibitem{Kurhe2011} Kurhe DN, Dagade DH, Jadhav JP, Govindwar SP, Patil KJ.  {Thermodynamic studies of amino acid-denaturant interactions in   aqueous solutions at 298.15 K}.  {J Solution Chem} 2011; 40:1596--1617.
\bibitem{Guinn2011} Guinn EJ, Pegram LM, Capp MW, Pollock MN, Record Jr. MT.  {Quantifying why urea is a protein denaturant, whereas glycine   betaine is a protein stabilizer}.  {Proc Natl Acad Sci USA} 2011; 108:16932--16937.
\bibitem{Diehl2013} Diehl RC, Guinn EJ, Capp MW, Tsodikov OV, Record Jr. MT.  {Quantifying additive interactions of the osmolyte proline with   individual functional groups of Proteins: comparisons with urea and glycine   betaine, interpretation of m-Values.}  {Biochemistry} 2013; 52:5997--6010.
\bibitem{Lin1994} Lin T, Timasheff SN.  {Why do some organisms use a urea-methylamine mixture as osmolyte?   Thermodynamic compensation of urea and trimethylamine N-oxide interactions   with protein}.  {Biochemistry} 1994; 33:12695--12701.
\bibitem{Timasheff1993} Timasheff SN.  {The control of protein stability and association by weak   interactions with water: how do solvents affect these processes?}  {Annu Rev Biophys Biomol Struct} 1993;   22:67--97.
\bibitem{Auton2007c} Auton M, Bolen DW.  {Application of the transfer model to understand how naturally   occurring osmolytes affect protein stability}.  {Methods Enzymol} 2007; 428:397--418.
\bibitem{Auton2004} Auton M, Bolen DW.  {Additive transfer free energies of the peptide backbone unit that   are independent of the model compound and the choice of concentration scale}.  {Biochemistry} 2004; 43:1329--1342.
\bibitem{Ben-Naim1987} Ben-Naim A.  {Solvation thermodynamics}. New York: Plenum 
Press; 1987. 246 p.
\bibitem{Ben-Naim2006} Ben-Naim A. {Molecular theory of solutions}. New York: 
Oxford University Press; 2006. 380 p.
\bibitem{Auton2007a} Auton M, Holthauzen LMF, Bolen DW.  {Anatomy of energetic 
changes accompanying urea-induced protein   denaturation}.  {Proc Natl Acad Sci 
USA} 2007; 104:15317--15322.
\bibitem{Record2013} Record Jr. MT, Guinn E, Pegram L, Capp M.  {Introductory lecture: Interpreting and predicting Hofmeister salt   ion and solute effects on biopolymer and model processes using the solute   partitioning model}.  {Faraday Discuss} 2013; 160:9--44.
\bibitem{Guinn2013a} Guinn EJ, Schwinefus JJ, Cha HK, McDevitt JL, Merker WE, Ritzer R, Muth GW, Engelsgjerd SW, Mangold KE, Thompson PJ, Kerins MJ, Record Jr. MT.  {Quantifying functional group interactions that determine urea   effects on nucleic acid helix formation.}  {J Am Chem Soc} 2013; 135:5828--5838.
\bibitem{Auton2005} Auton M, Bolen DW.  {Predicting the energetics of 
osmolyte-induced protein   folding/unfolding}.  {Proc Natl Acad Sci USA} 2005; 
102:15065--15068.
\end{thebibliography}
\end{document}